# Reconstructions of refractive index tomograms via discrete algebraic reconstruction technique

**MOOSUNG LEE,[1,2] SEUNGWOO SHIN,[1,2] AND YONGKEUN PARK[1,2,3*]**

[1]Department of Physics, Korea Advanced Institute of Science and Technology (KAIST), Daejeon, 34141, South Korea
[2]KI for Health Science and Technology, KAIST, Daejeon 34141, Republic of Korea
[3]Tomocube, Inc., Daejeon 34051, Republic of Korea
*yk.park@kaist.ac.kr

**Optical diffraction tomography (ODT) provides three-dimensional refractive index (RI) tomograms of a transparent microscopic object. However, because of the finite numerical aperture of objective lenses, ODT has the limited access to diffracted light and suffers from poor spatial resolution, particularly along the axial direction. To overcome the limitation of the quality of RI tomography, we present an algorithm that accurately reconstructs RI tomography using preliminary information that the RI values of a specimen are discrete and uniform. Through simulations and experiments on various samples, including microspheres, red blood cells, and water droplets, we show that the proposed method can precisely reconstruct RI tomograms of samples with discrete and uniform RI values in the presence of the missing information and noise.**

## 1. Introduction

Measuring three-dimensional (3D) refractive index (RI) provides invaluable information about a sample because it enables quantitative and label-free characterization of transparent microscopic objects such as biological cells and tissues. In general, optical diffraction tomography (ODT) or 3D quantitative phase imaging (QPI) techniques measure multiple 2D holograms of a sample with various illumination angles, from which a 3D RI tomogram of the sample is reconstructed [1-3]. Although the principle of ODT was first demonstrated shortly after the birth of the laser [4], the field has rapidly advanced since 2000s when 3D RI tomograms began to be used for the study of biological samples [5-7], largely owing to its capability of label-free live-cell imaging and quantitative characterization. Although ODT has opened the door for many biological, biochemical, and industrial applications to be studied with 3D RI tomograms [8-12], it suffers from anisotropic resolution due to the missing cone problem [Fig. 1].

When ODT measures multiple 2D holograms of a sample at various illumination angles to reconstruct its RI distribution, an optical imaging system cannot collect side-scattered signals due to the finite numerical aperture (NA) [Fig. 1(a)]. The uncollected signals correspond to the missing information about the scattering potential in Fourier space, also known as the missing cone problem [Fig. 1(b)]. As a result, a reconstructed 3-D RI map has poor axial resolution and suffers from artifacts around the sample boundary with imprecise RI distribution [Fig. 1(c)].

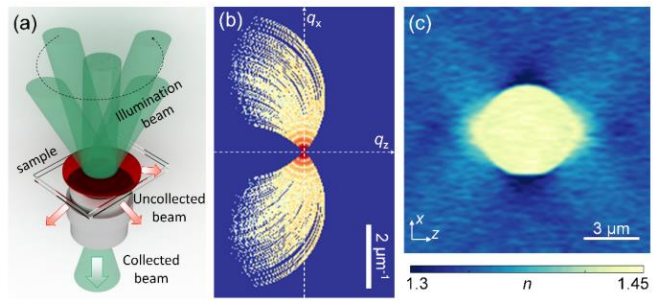

Fig. 1. Missing cone problem. (a) The finite numerical aperture of an imaging system results in limited access to the scattering signal from a sample. (b) Simulated scattering potential represented in Fourier space. (c) The RI map of a simulated silica bead retrieved from (b).

In order to circumvent the missing cone problem and enhance the reconstruction performance of ODT, several methods have been suggested. For example, sample rotation techniques have been developed using microfluidic channels [13, 14], fiber needles [15], and optical tweezers [16]. These methods, however, require special setups and have difficulty in handling soft biological samples. Illumination rotation methods have also been presented using spatial light modulators, including galvanometric mirrors [15, 17], liquid crystal spatial light modulators [18-20], and digital micromirror devices [21, 22]. Still, however, the techniques suffer from the missing cone and the resultant anisotropic resolution. Alternatively, various reconstruction algorithms utilizing appropriate constraints and regularizations have been developed. A comparative study of algorithms, for example, showed that total variation minimization algorithm significantly reduces noises in reconstructed RI tomograms [20]. Nevertheless, the algorithm does not enhance the axial resolution. As complementary methods, multiple light scattering has been concerned using beam propagation methods [23, 24], wave propagation methods [23], and the Lippmann-Schwinger equation [24-26]. However, these techniques require heavy computations due to the optimization of multiple regularization parameters.

Here, we present a new algorithm that accurately reconstructs a 3D RI tomogram of a sample with discrete and uniform RI values, exploiting priori information about RI features. Inspired by the discrete algebraic reconstruction technique (DART) used in X-ray computed tomography (CT) [27-29], this presented algorithm, named ODT-DART, successfully reconstructs a discrete 3D RI tomogram with small computational cost, noise tolerance, and high accuracy. We first explain the principle and formulate the algorithm, and then verify that ODT-DART provides a precise and robust reconstruction of discrete RI tomograms in the presence of missing cone problems and noise in measurements. We also present the experimental result of various samples, including colloids, red blood cells, and microdroplets of water, and discuss the potentials and limitations of the present method.

## 2. Principles of ODT-DART

Figure 2 illustrates the flowchart of ODT-DART. The whole sequential steps of ODT-DART consist of (step 1) the first iterations of Gerchberg-Papoulis (GP) algorithm, (step 2) intermediate discretization, (step 3) the second iterations of GP, (step 4) final discretization, and (step 5) voxel error correction after iterations of steps 1–4. Each procedure works as follows:

(step 1) the first iteration of GP algorithm: To reduce the RI artifact caused by the missing cone, the GP algorithm is conducted with physical constraints to both image and Fourier spaces of a given tomogram [30, 31]. In the image space, constraints of non-negativity on RI are applied: $\text{Re}(n) \geq n_m$, where $n_m$ is the background medium. After applying the Fourier transform, the Fourier space is updated with the experimentally measured scattering potential $F^{(0)}(\mathbf{q})$, where $\mathbf{q}$ is a wave vector. In Fourier space, the scattering potential is limited within the range of far-field bandwidth ($|\mathbf{q}| \leq q_{max} = 4\pi n_m / \lambda$), where $\lambda$ is the wavelength of the illumination. After the inverse Fourier transform, the scattering potential is converted into the updated distribution of complex RI values in image space. These procedures are repeated 25 times for stable convergence of the GP algorithm.

(step 2) Intermediate discretization: A threshold function for discrete RI tomography is applied. We define the heuristic threshold function, $T(n, R)$, where n is the input RI and $R$ is the set of the background RI and $j$ possible discrete RI values of a sample, i.e. $R = [n_m, n_1, \ldots, n_j]$. In experiments, we estimated the number and the values of RI of a sample from the manufacturers' information (e.g. silica beads) or references (e.g. water and red blood cells) [32, 33]. The thresholds of the function are defined as the averages of two consecutive RI values,

$$T(n,R) = \begin{cases} n_m, & \text{if } n < (n_m + n_1)/2, \\ n_{i, 2 \leq i < j}, & \text{if } (n_{i-1} + n_i)/2 \leq n < (n_i + n_{i+1})/2, \\ n_j, & \text{if } n \geq (n_{j-1} + n_i)/2. \end{cases} \quad (1)$$

Note that, in order to avoid the rough artifacts on the discrete boundaries, step 2 employed the more number of discrete RI values than the priori information about the sample. For example, the set of R can be oversampled by a factor of 2–3, i.e. $R' = [n_m, (2n_m + n_1)/3, (n_m + 2n_1)/3, n_1, \ldots, n_j]$.

(step 3) the second iteration of GP algorithm: In order to smooth the RI tomogram and update the experimentally measured scattering potential, the GP algorithm is applied to the result of step 2 five times.

(step 4) Final discretization: The threshold function with the $j$ possible discrete RI values is applied to the result of step 3. To retrieve the exact RI values of unspecific discrete objects, a regularized least square method is applied. The following error function is formulated:

$$E(\psi) = \frac{1}{p} \sum_{\{\mathbf{q} \in F^{(0)}(\mathbf{q}) \neq 0\}} \left| F^{(0)}(\mathbf{q}) - \sum_{t=1}^{j} \rho_t B_t(\mathbf{q}) \right|^2 + \varepsilon \sum_{t=1}^{j} |\rho_t|^2 \quad (2)$$

where $\psi$ is the set of discrete values, $\psi = [\rho_1, \ldots, \rho_j]$, $\rho_t = -(2\pi n_m/\lambda)2[(n_t/n_m)^2 -1]$, $B_t(\mathbf{q})$ is the Fourier transform of the tth sample region having the RI value of $n_t$, $p$ is the number of the non-zero voxels in $F^{(0)}(\mathbf{q})$, and $\varepsilon$ is the regularization parameter ($\varepsilon = 0.005$ for the results shown below).

The minimization of this error function can be achieved by solving $\partial E/\partial \psi = 0$ (see Appendix 1). Then, the discrete RI values in the sample region are updated by solving the following equation:

$$-\left(\frac{2\pi n_m}{\lambda}\right)^2 \left[(n_i^{new}/n_m)^2 - 1\right] = \left(\varepsilon p I + B^{\dagger} B\right)^{-1} \quad (3)$$

where $i$ is the index ranging from 1 to $j$, $n_i^{new}$ is the updated RI in the $i$th sample region, $B = [\mathbf{b_1}, ..., \mathbf{b_j}]$ is a $p \times j$ matrix, $I$ is a $j \times j$ identity matrix, and $\mathbf{b}_i$ and $\mathbf{f}$ are $p \times 1$ vectors representing the voxels of $F^{(0)}(\mathbf{q} \in F^{(0)}(\mathbf{q}) \neq 0)$ and $B_t(\mathbf{q} \in F^{(0)}(\mathbf{q}) \neq 0)$ (the pink areas in Fig. 2).

(step 5) Voxel error correction: After three iterations of steps 1–4, the reconstructed discrete RI tomogram has a few voxels along edges showing highly discontinuous pixelated artifacts. Because the samples tested in our demonstration were larger than the diffraction limit, we defined the voxel errors as discrete regions smaller than the diffraction limited spot size of the ODT system. Then, to minimize the voxel errors, the RI values of these voxel artifacts are matched with those of their neighboring voxels. To resolve smaller samples, this step can be omitted or use a smaller threshold volume. The scattering potential in the Fourier space during the process shows that each step significantly fills up the missing cone. Further details about the ODT-DART algorithm can be found in the pseudocode in Appendix 1.

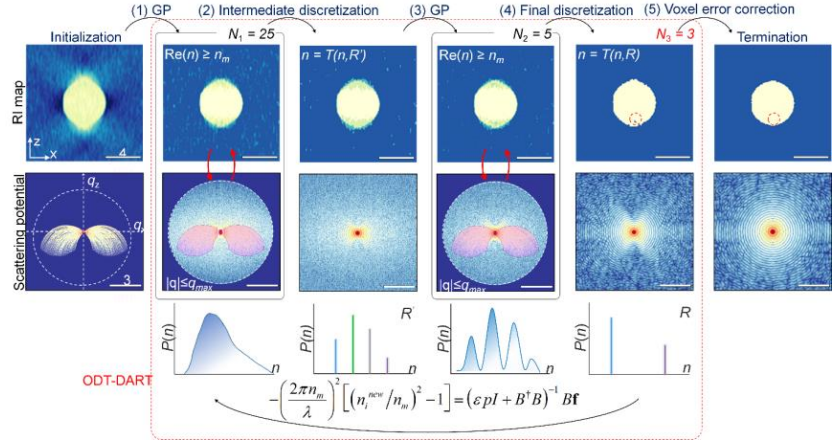

Fig. 2. Schematic diagram of the ODT-DART algorithm. Step 1: 25 iterations of GP algorithm. Step 2: Intermediate discretization with sufficient number of discrete RI values. Step 3: 5 iterations of GP algorithm. Step 4: Final discretization with modified least squares Step 5: Voxel error correction after 3 iterations of Steps 1-4.

## 3. Simulation

In order to validate the reconstruction performance of ODT-DART, we performed numerical simulations with discrete phantoms and compared the performance of the present method with a conventional method (GP algorithm, 100 iterations) [Fig. 3]. To compare a realistic ODT system with coherent speckle noise, we also tested the reconstruction performances in the absence and the presence of white Gaussian noises (the standard deviation of 0.1, -10 dB), respectively.

We first tested a phantom representing a 5-μm-diameter microsphere ($n_s = 1.461$) suspended in water ($n_m = 1.336$) [Figs. 3(a)-(b)]. The 3-D RI tomograms reconstructed with GP well agree with the ground trough, especially along the focal plane. However, the RI profiles exhibit stretching along axial directions and fluctuations in RI values. This error becomes more significant in the presence of additional noise. In contrast, the RI tomograms reconstructed with the present method showed good agreements with the ground truth in terms of shapes and RI values in both the lateral and axial directions. Even in the presence of noise, the result with ODT-DART provides clear reconstructions of an object.

For the quantitative analysis, the accuracy was calculated as the mean squared error between the RI tomograms of the ground truth and the reconstructed. The results with ODT-DART show mean squared errors of 0.494 and 0.859 ppm for the absence and the presence of noise, respectively, whereas the results with GP show 2.03 and 19.17 ppm, respectively. The cross-sectional images of the RI maps demonstrate that the results with ODT-DART are in good agreements with the original sample information in terms of both the morphology and RI values [Fig. 3(c)].

To further validate the capability of ODT-DART for reconstructing tomograms with multiple RI values, we conducted similar numerical simulations with an object with encapsulating geometry. The phantom was aqueous suspension ($n_m = 1.336$) of an 8-μm-diameter microsphere ($n_s = 1.4$) encapsulating a 3-μm-diameter microsphere ($n_s = 1.461$) at $z = 1$ μm [Figs. 3(d)−(f)].

The 3D RI reconstruction obtained with the conventional GP presents artifacts, especially along the axial direction, caused by the limited NA [Fig. 3(d)]. The presence of the noise severely deteriorates the reconstructed image quality. In contrast, the results obtained with ODT-DART exhibit clear and precise reconstructions of the original shape, regardless of the presence of noise [Fig. 3(e)]. The results with ODT-DART present mean squared error of 0.444 and 1.45 ppm for the absence and the presence of noise, respectively, whereas the results with GP show 2.01 and 20.59 ppm, respectively. The shapes of both the internal and external boundaries of the sample are reconstructed and the RI values are well consistent with the ground truth [Figs. 3(f)].

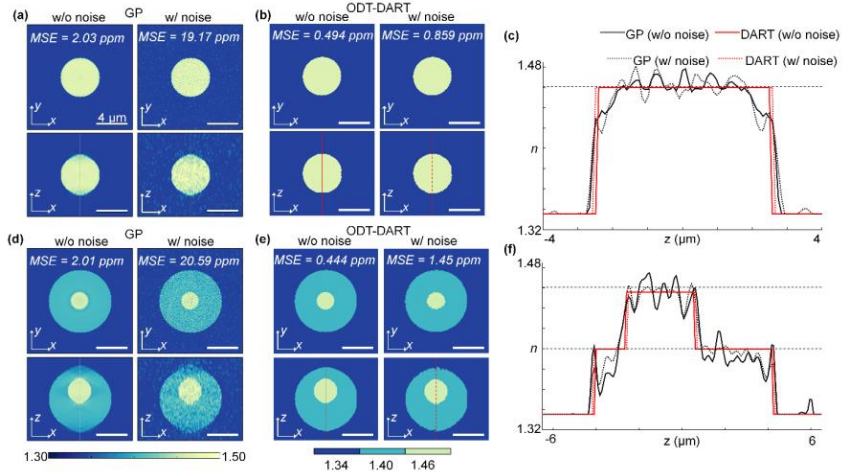

Fig. 3. The numerical simulations and performances of ODT-DART in comparison to GP algorithm. (a-b) cross-sectional images of the RI maps of 5-μm-diameter microspheres ($n_s$ = 1.461) reconstructed by (a) GP and (b) ODT-DART. (c) The axial RI profile along the center of the reconstructed tomograms in (a-b). (d-e) cross-sectional images of the RI maps of 8-μm-diameter microsphere ($n_s$ = 1.4) encapsulating a 3-μm-diameter microsphere ($n_s$ = 1.461) reconstructed by (d) GP and (e) ODT-DART. The background medium was set to have RI of $n_m$ = 1.336. The left (right) images are the reconstructed tomograms in the absence (presence) of the white Gaussian noises (-10 dB) in the simulated sample fields. (f) The axial RI profile along the center of the reconstructed tomograms in (d-e).

## 4. Experiment

The performance of ODT-DART is validated with the experimental data. We applied ODT-DART algorithms to reconstruct 3D RI tomograms of various soft material and biological samples including a colloidal suspension, a healthy human red blood cell (RBC) and a RBC parasitized by Plasmodium falciparum (Pf-RBC), and a water droplet. The details of the experimental procedure are explained in Appendix 2.

### 4.1 Imaging a silica microsphere

To test the reconstruction accuracy in a simple case, we reconstructed the tomogram of a silica microsphere (5 μm diameter, 44054-5ML-F, Sigma-Aldrich Inc., USA) suspended in deionized water [Figs. 4(a)−(c)]. As in simulations, the RI tomogram reconstructed with GP suffers from noise in both the sample and background area [Fig. 4(a)]. However, ODT-DART clearly reconstructs the overall morphology of the sample without image noise [Figs. 4(b)−(c)]. The retrieved RI value (1.427) and the calculated volume (65.32 fL) are in good agreement with the sample specification.

### 4.2 Imaging red blood cells

To validate the applicability to biological samples, a human RBC is reconstructed with ODT-DART. According to standard protocol, a RBC extracted from a healthy individual was diluted with a Dulbecco's phosphate-buffered saline (DPBS) solution before measurement.

The results are shown in Figs. 4(d)−(f). The RI tomogram of a RBC reconstructed by GP exhibit significant pixelated noise and blurred RI distribution near its thin center due to the missing cone and noise [32]. On the contrary, the results reconstructed with ODT-DART clearly show the characteristic biconcave structure of a RBC whose minimum thickness was 925.7 nm. From the reconstructed 3D RI maps of the RBC, the volume was measured as 103.30 fL. The Hb concentration of the RBC was calculated from the cytoplasmic RI value ($n_s$ = 1.396) based on Ref. [34], which was 30.66 g/dL. The total amount of Hb was 31.67 pg. These values were in agreement with physiological levels of RBCs [35].

To further verify with experimental data with multiple RI values, we applied ODT-DART to reconstruct the 3D RI map of a Pf-RBC from the raw data obtained in Ref. [33]. As in Figs. 4(g)−(i), the reconstructed 3D RI tomogram clearly visualizes the Pf-RBC in a trophozoite stage of the intraerythrocytic cycle of malaria infection [36]. In the reconstructed 3D RI map, the shape of a hemozoin, a byproduct of Hb digestion by malaria parasites, is clearly visualized. Also, the alterations of the shapes of the host cell membrane and the decrease in the cytoplasmic RI, implies the loss of Hb. In both GP and ODT-DART, the hemozoin structures were clearly visualized since they have high RI values ($n$ = 1.447). However, the result obtained with GP does not clearly visualize the structures of the cell membrane, whereas the present method successfully reconstructs the membrane of the host cell [Figs. 4(h)−(i)]. The characteristic structures in the Pf-RBC such as "hole" in cytoplasm and irregular membrane shapes imply the parasitic vacuoles and alterations in the host cell membrane, considering the pathophysiology of malaria infection [36, 37]. Nonetheless, further confirmation of these structures may require correlative imaging with fluorescence [20, 38].

The accurate reconstruction of the 3D RI tomograms of Pf-RBC leads to better quantitative analysis of the cells. For example, the cell volume, Hb concentration, and total Hb contents of the PF-RBC were retrieved as 85.43 fL, 29.45 g/dL, and 25.16 pg, respectively. The decrease of the Hb concentration and Hb contents in the infected cell is consistent with the fact that a malaria parasite metabolizes cytoplasmic Hb and generates hemozoin [36].

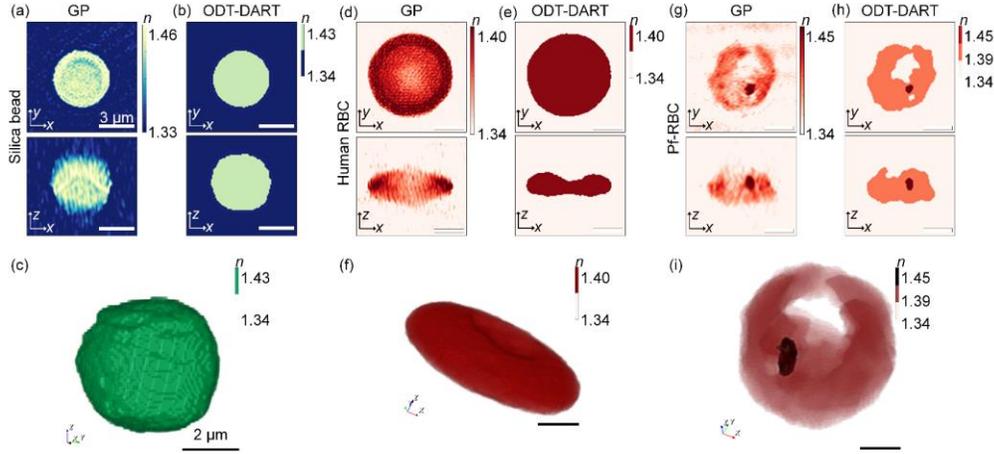

Fig. 4. Experimental results with a silica microsphere, a RBC and a Pf-RBC. (a-b) cross section images of the RI maps of a 5-μm-diameter silica microsphere suspended in water reconstructed by (a) GP and (b) ODT-DART. (c) 3D-rendered tomogram of (b). (d-e) Cross sections of a human RBC reconstructed by (c) GP and (d) ODT-DART. (f) 3D-rendered tomogram of (e). (g-h) Cross sections of a Pf-RBC in the trophozoite stage reconstructed by (g) GP and (h) ODT-DART. (i) 3D-rendered tomogram of (h).

*4.3 Quantitative 3D imaging dynamics of an evaporating water drop*

To further test the applicability of ODT-DART for time-lapse 3D imaging, the evaporation dynamics of a microscopic water droplet were measured and reconstructed [Fig. 5]. To prepare a microdroplet, distilled water was sprayed onto a coverslip; then the coverslip was sealed with another coverslip to avoid rapid evaporation. Then, 60 tomograms of an evaporating droplet were measured for 1 hour.

Figures 5(a)−(b) illustrate the reconstructed tomograms of the initial droplet, obtained by GP without and with the geometric constraint. The area corresponding to the bottom coverslip was excluded in the RI reconstruction procedure. The tomogram reconstructed by GP generally exhibits a swollen shape and lower RI than the actual RI ($n_s$ = 1.336), especially for such a thin sample. With the geometric constraints, the result shows flat bottom and clearer outlines [Fig. 5(b)]. However, the results still exhibit gradual variations of RI along the axial direction due to the missing cone problem.

In contrast, ODT-DART in combination with the geometric constraint precisely reconstructed the microdroplet in 3D [Fig. 5(c)]. The overall shape of the microdroplet was clearly reconstructed, without blurring along the axial direction. The mean RI value of the droplet was 1.336 ± 0.005 (averaged from 60 tomograms) and well agrees with the RI value of water.

The precise reconstruction capability of ODT-DART enables quantitative analysis of the evaporation dynamics of a water droplet in a confined system [Figs. 5(d)−(e) and Visualization 1]. The volume of the water droplet was calculated from the RI tomograms, which showed the non-linear decreases as a function of time and reached a dynamic equilibrium after 30 min [Fig. 5(d)]. This process was different from the evaporation in an open system in which the droplet volume decreases linearly [37]. Furthermore, the dynamics of 3D RI maps show that the center of mass of the droplet moved towards the pinned structure [the black arrow in Fig. 5(e)], and stopped when the pin disappeared, indicating a typical slip-stick motion in a rough surface. This asymmetric displacement in a confined system has more implications in soft matter physics, which can now be accessible with the present method.

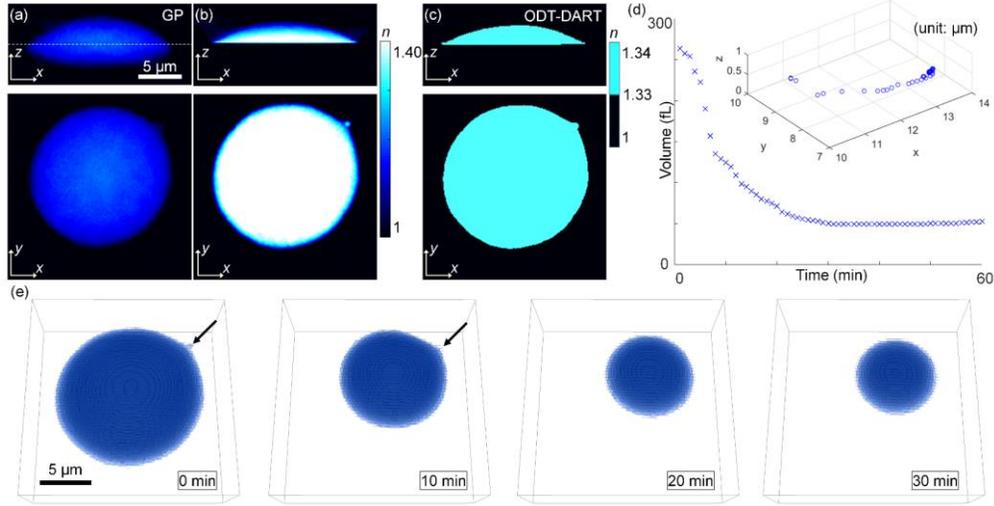

Fig. 5. (a-b) Cross sections of the 3-D RI map of a water droplet reconstructed by (a) GP, (b) GP with the geometric constraint, and (c) ODT-DART with the geometric constraint. (d) Volumes of an evaporating water drop as a function of time. Inlet: 3D tracking of the center-of-mass of the water drop in 3D. (e) Visualizations of evaporating water drop over one hour (see Visualization 1).

## 5. Discussions

In summary, we present a method to reconstruct 3D RI tomograms of samples with discrete RI distributions. Exploiting the prior knowledge about RI discreteness, the ODT-DART algorithm provides reconstructions of RI maps of a sample with high precision and resolution. The present method overcomes the mission cone problem because the information which was not collected by the finite NA of an imaging system is filled with prior information about RI discreteness. The capability of ODT-DART was demonstrated with various types of samples, including microspheres, RBCs, and liquid droplets.

Motivated from the DART algorithms in X-ray CT [29, 36, 39], ODT-DART enables high-resolution and high-precision tomographic reconstructions of 3D RI maps. The current approach is different from a recent work where an approximate knowledge of the sample permittivity was used to enhance the spatial resolution in the reconstruction [40]. For example, ODT-DART does not require the knowledge of the sample permittivity; it only requires the discreteness and the number of the discrete RI values.

The presented results suggest that, compared with the previous GP algorithm, ODT-DART can improve the reconstruction accuracy using appropriate constraints based on the prior sample information. However, in principle, our method does not guarantee the improved spatial resolution to reconstruct objects smaller than the diffraction limit (see Appendix 3). To overcome the diffraction-limited resolution, further elaborations are necessary to consider various aspects, such as parameter optimizations and multiple light scattering [40, 41].

One of the limitations of the present approach is the prior knowledge and requirement of discreteness of RI distributions of a sample. Also, the present approach has limited performances for samples with multiple discrete boundaries (see Appendix 3). Nevertheless, most of the current applications of ODT deal with samples with a few number of discrete RI values. In addition, the speed of the reconstruction should also be enhanced. In the present work, the typical reconstruction time was three minutes or sooner for a tomogram of 300 × 300 × 300 voxels when operated on a desktop computer (See Appendix 4). However, the reconstruction speed can be dramatically enhanced when a graphical processor unit is utilized [40].

From a technical point of view, the present method can further be expanded to spectroscopic tomography by using a wavelength-scanning illumination [42]. In addition, the reduction of speckle noise using incoherency of illumination may further improve the quality of 3D RI tomography [43].

ODT-DART algorithm is robust, accurate and potentially adaptable to a wide variety of research fields. It thus has the potential for direct applications including biophysics, material sciences, cell biology, and infectious diseases. Although only the real part of RI tomograms are shown in the results, ODT-DART algorithm in principle reconstructs the complex RI tomograms, consisting of both the absorption and phase relay part. This may also open the door to other applications where absorptive parts of a sample are important, such as hemozoin in Pf-RBC (see Appendix 5). In addition, the instrumentation for measuring 3D RI maps has been recently commercialized [44], which will expedite the application of the present algorithm.

## Appendix

*1. Detailed algorithm formulation*

Table 1 depicts the pseudocode of ODT-DART in MATLAB. We initialize ODT-DART with the measured scattering potential, $F^{(0)}(\mathbf{q})$, and a set of the background medium and initially guessed RIs of a sample, $R^{(0)} = (n_m, n_1^{(0)}, \ldots, n_j^{(0)})$. The spatial frequency, $k$, is $2\pi n_m/\lambda$ ($n_m/\lambda$ in MATLAB convention). $p$ counts the nonzero voxels in $F^{(0)}(\mathbf{q})$. The details about each of five steps are illustrated below:

(Step 1) Gerchberg-Papoulis (GP) algorithm: We applied GP to reduce the artifacts due to the missing cone. First, we applied the non-negativity constraint on RI ($\text{Re}(n) \geq n_m$). The modified complex RI map was translated to the scattering potential as $f(\mathbf{x}) = -k^2[(n(\mathbf{x})/n_m)^2 - 1]$, and numerically Fourier-transformed by fast Fourier transform (FFT).

In Fourier space, we first matched the updated potential with $F^{(0)}(\mathbf{q})$. We then limited the bandwidth of the scattering potential to the far-field limit ($|\mathbf{q}| \leq 2k$). We could retrieve the corrected RI map using inverse FFT and algebra. This completed one loop of GP. We took 25 iterations in this step.

(Step 2) Intermediate discretization: To circumvent the heavy fluctuations of sample boundaries, we employed intermediate discretization with more RI elements, $R' = (n_m, (2n_m + n_1)/3, (n_m + 2n_1)/3, n_1, \ldots, n_j)$. We discretized the RI map using the RI threshold function defined in Eq. (1) of the main text, $T(n(\mathbf{x}), R')$, where the threshold is simply the mean of two consecutive RI elements.

(Step 3) GP: In this step, we applied five iterations of GP with some modifications. During iterations, we fixed 85 % of RI maps having the RI value in the set of sample RI values, R. This enabled to update the sample boundaries until convergence while maintaining the correct sample regions. The modified RI map was smoothed by Gaussian convolution. In the last loop when $l = 3$, we did not conduct smoothing in order to maintain the sharpness of the RI map.

(Step 4) Final discretization: We discretized the modified RI map with the threshold function, $T(n(\mathbf{x}), R)$. We then updated the RI values of the tomogram by regularized least square, as in eq. (2) of the main text. At the minimum, the error function satisfies the following relation:

$$\frac{\partial E(\psi)}{\partial \rho_k^*} = 0 = -\frac{2}{p} \sum_{\{\mathbf{q} \in F^{(0)}(\mathbf{q}) \neq 0\}} \left( F^{(0)}(\mathbf{q}) - \sum_{t=1}^{j} \rho_t B_t(\mathbf{q}) \right) B_k(\mathbf{q})^* + 2\varepsilon \sum_{t=1}^{j} \rho_t \quad (4)$$

, where $\psi$ is the set of discrete sample scattering values, $(\rho_1, \ldots, \rho_j)^T$, $B_t(\mathbf{q})$ is the Fourier transform of the sample region having RI value of $n_t$, and $\varepsilon$ is the regularization parameter that was set to be 0.005 in our demonstration. Rearranging the eq. (4), the following eq. (5) is obtained.

$$\sum_{t=1}^{j} \rho_t \left( \varepsilon p + \sum_{\{\mathbf{q} \in F^{(1)}(\mathbf{q}) \neq 0\}} B_k(\mathbf{q})^* B_t(\mathbf{q}) \right) = \sum_{\{\mathbf{q} \in F^{(1)}(\mathbf{q}) \neq 0\}} F^{(0)}(\mathbf{q}) B_k(\mathbf{q})^* \quad (5)$$

By vectorizing $F^{(0)}(\mathbf{q} \in F^{(0)}(\mathbf{q}) \neq 0)$ and $B_t(\mathbf{q} \in F^{(0)}(\mathbf{q}) \neq 0)$ as $\mathbf{f}$ and $\mathbf{b}_t$ and defining the matrix $\mathbf{B} = [\mathbf{b}_1, \ldots, \mathbf{b}_j]$, the equation (6) is simplified to the linear matrix equation.

$$\left( \varepsilon p I + B^\dagger B \right) \psi = B^\dagger \mathbf{f} \quad (6)$$

The inverse of the matrix on the left side leads to the eq. (2) in the main text.

To get the faster convergence, we applied the following minor treatments in this step. First, when we obtained $B_t(\mathbf{q})$, we smoothened the sample region in the image space to reflect the imperfect sample boundary of the trial solution. Second, we found that updating $R$ as the mean value of retrieved $R$ and $R^{(0)}$ showed faster convergence, given that the actual RI values of the sample are close to $R^{(0)}$.

**Table 1. MATLAB pseudo-code for ODT-DART**

```
g = 1; g' = 1; l = 1;
N₁ = 25; N₂ = 5; N₃ = 3;
k = nm/λ  % In MATLAB convention.
F⁽⁰⁾(q): Measured scattering potential
R⁽⁰⁾ = (nm, n₁⁽⁰⁾, …, nⱼ⁽⁰⁾); R = R⁽⁰⁾
p = Count(q ∈ F⁽⁰⁾(q) ≠ 0)
% Start ODT-DART
while l ≤ N₃
    % Start step 1: GP
        while g ≤ N₁
        % Non-negativity
            Re(ñ(x ∈ Re(ñ(x)) < nm)) = nm;
    % Start step 3: GP
        f⁽ˡ⁾(x) = -k²[(n⁽ˡ⁾(x)/nm)² -1]; F⁽ˡ⁾(x) = fftn(F⁽ˡ⁾(x));
        x' ∈ [n⁽ˡ⁾(x') R⁽ʲ⁾ in 85% probability];
        n' = n⁽ˡ⁾(x');
        while g' ≤ N₂
            ñ⁽ˡ⁾(x')=n';
            … % GP with ñ⁽ˡ⁾(x')
        end
        if l ≠ N₃: n(x) = smooth3(n(x)); end
    % End step 3
    % Start step 4: Final discretization
        n(x) = T(Re[ñ(x)], R);
```

```
    f(x) = -k²[(ñ(x)/nₘ)² -1];                    f  = F(q)[F⁽⁰⁾(q) ≠ 0];   % vectorize
    F(q) = fftn(F(x));                             bᵢ = fftn(n(x) = nᵢ)[F⁽⁰⁾(q) ≠ 0]; B = [b₁ … bⱼ];
    g = g + 1;
% Matching measured potential                     ψ = (B†B + εpI)⁻¹ B†f
    F(q) = F(q ∈ F⁽⁰⁾(q) = 0) + F(q);             (n₁, …, nⱼ) = Re[nₘ√[1- (ψ/k)²]];
% Constraint in reciprocal space                  R= (nₘ, (n₁⁽⁰⁾+ n₁)/2, …);
    F(q) = F(q) × (|q| ≤ 2k);                     l = l+ 1;
    f(x) = ifftn(F(q));                         end
    ñ(x) = nₘ√[1- (f(x)/k)²];                   % End step 4
end % End step 1                                % Step 5: Voxel error correction
% Step 2: Intermediate discretization            Threshold = V_PSF % Can be set as smaller values.
R' = (nₘ, (2nₘ+ n₁)/3, (nₘ+ 2n₁)/3, …)          n(x) = Mostneighbor(n(x),Threshold)
n(x) = T(Re[ñ(x)], R');                         Return n(x)
```

(Step 5) Voxel error correction: After three iterations of steps 1−4, the retrieved RI map had remaining voxel errors. We removed these errors based on *bwconncomp* function, which finds the connected components in our discrete RI map. Because the reconstructed samples in our demonstrations were far larger than the diffraction limit, we defined the voxel errors as discrete regions that have smaller volumes than the volume of the point spread function, where the lateral and axial resolution, *Δx* and *Δz*, are defined as follows [45]:

$$\Delta x = \frac{\lambda}{n_m \sin\theta_{scan} + NA_{obj}} \tag{7}$$

$$\Delta z = \frac{2\lambda}{2n_m - \sqrt{n_m^2 - n_m^2 \sin^2\theta_{scan}} - \text{Re}\left[\sqrt{n_m^2 - NA_{obj}^2}\right]} \tag{8}$$

These defined artifacts were matched to the RI value that the majority of the neighbor voxels have. Note that, for smaller samples, we may reduce the size of the threshold volume or omit the process.

We operated a desktop computer (Intel Core i5-6600, 3.3 GHz, 16 GB RAM) to process the tomographic reconstruction. We employed MATLAB R2014b (Mathworks Inc.) to formulate the algorithm for ODT-DART.

*2. Detailed experimental setup*

To demonstrate the applications of ODT-DART in experiments, we imaged discrete RI objects using an off-axis Mach-Zehnder interferometer [Fig. 6(a)] coupled with a high-NA condenser objective lens. We scanned incident illuminations using a beam modulator unit [Fig. 6(b)]. The employed device was either a digital micromirror device or a pair of galvanometric mirrors. As in Fig. 6(c), we retrieved the sample fields from the recorded fringe patterns using Fourier transform [44]. We converted the measured sample fields to the scattering potential using Rytov approximation and Fourier diffraction theorem [46, 47]. We validate the reconstruction performances of ODT-DART in experiments on various soft matters including a colloidal particle, red blood cells, and microscopic water droplets.

We employed three different optical configurations for acquiring multiple holograms at different illumination angles. To acquire the tomograms of a silica microsphere and a healthy human red blood cell, we employed the optical setup illustrated in ref. [21], in which a high-NA condenser lens (UPLASAPO 60XW, Olympus Inc., Japan) collimated the first diffraction-order from a digital micro-mirror device (DLP LightCrafter 3000, Texas Instruments, Inc., USA) to change the incident angle of illuminations in a spiral pattern. The objective lens (UPLASAPO 60XO, Olympus Inc., Japan) and the condenser lens supported $NA_{obj}$ = 1.42 and $n_m\sin\theta_{scan}$ = 0.9 respectively. We employed a fiber coupler (TW560R2F2, Thorlabs, Inc., USA) to separate the green beam from a diode-pumped solid state laser with a single longitudinal mode ($\lambda$ = 532 nm, 50 mW, Cobolt Co., Solna, Sweden) into the sample arm and reference arm. Our CCD camera (FL3-U3-13Y3M-C, FLIR Systems, Inc., USA) recorded the fringe patterns with a magnification of 60.

To acquire the tomograms of a water droplet, we utilized the optical configuration illustrated in ref. [48]. A high-NA condenser lens (UPLASAPO 60XW, Olympus Inc., Japan) collimated the deflected beam from a pair of galvanometric mirrors (GVS012/M, Thorlabs Inc., USA). The illumination angles were scanned in a spiral pattern with $n_m\sin\theta_{scan}$ = 0.7. A high-NA objective lens (UPLASAPO 60XO, Olympus Inc., Japan) collected the scattered field, which was recorded by a sCMOS camera (Neo sCMOS, Andor Inc., UK) with a magnification of 400. The illumination source was the same as that of the first setup.

To acquire the tomogram of a malaria-infected red blood cell in a trophozoite stage, we employed the experimental configuration illustrated in Ref. [49]. We employed a pair of galvanometric mirrors (GVS012, Thorlabs Inc., USA) to scan the illumination angles linearly with $n_m\sin\theta_{scan}$ = 1.23. The condenser and objective lenses (UPLSAPO, 100×, Olympus Inc., USA) provided the high resolution of the images acquired by a high-speed CMOS camera (1024 PCI, Photron USA Inc., USA) with a magnification of 210. The illumination source was a He-Ne laser (λ = 633 nm, 10 mW, Throlabs, USA).

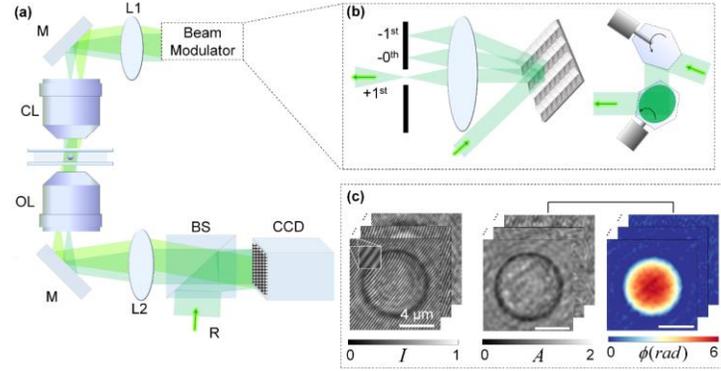

Fig. 6. (a) Schematic optical setup for reconstructing RI tomograms. M: mirror, L: convex lens, CL: condenser lens, OL: objective lens, BS: beam splitter, and R: reference beam. (b) A beam modulator unit. Left: 1st diffraction order from a digital micro-mirror device. Right: A pair of galvanometric mirrors. (c) Measured off-axis holograms at multiple angles and retrieved amplitude and phase images.

## *3. Performance limits of ODT-DART*

To investigate the applicability of the present method, we conducted various additional simulations in which ODT-DART has limited reconstruction performances [Fig. 7]. To study the resolution limit of ODT-DART, we first tested a 200-nm-diameter silica bead, which is smaller than the lateral and axial resolution of 205 and 510 nm, respectively [Fig. 7(a)]. The regularization parameter was as $\varepsilon = 10^{-6}$. Both GP and ODT-DART work for this sample, but the reconstructed tomograms showed elongations in the axial direction [Fig. 7(a)]. The MSE results were 4.35 and 6.33 ppm for GP and ODT-DART, respectively. The lateral and axial profile clearly show that ODT-DART has laterally squeezed artifacts, largely owing to the imperfect discretization [Fig. 7(b)]. The lateral and axial lengths of the reconstructed tomograms by GP were 249 and 369 nm, respectively, whereas those by ODT-DART were 149 and 254 nm, respectively.

Then, we tested a 3D Shepp-Logan phantom [50] which has five discrete RI values [Figs. 7(c)–(d)]. Cross sectional images of the reconstructed RI tomograms show that both GP and ODT-DART have artifacts, in particular around the axial ends. In the absence of noises, the MSE results were 4.02 ppm and 2.99 ppm for ODT-DART and GP (2.99 ppm), respectively. However, ODT-DART still shows the robust reconstruction of large discrete lobes. The MSEs of GP and ODT-DART were 6.53 and 56.64 ppm, respectively.

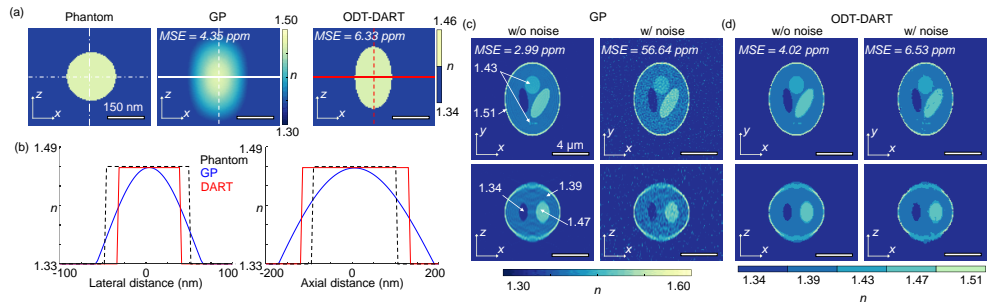

Fig. 7. Limited performances of ODT-DART in simulations. (a) Comparison of tomograms of a 200-nm-diameter silica bead reconstructed by GP and ODT-DART. (b) The lateral and axial profile of phantom and reconstructed RI tomograms. (c-d) Reconstruction performance for 3D Shepp-Logan phantom with RI values of 1.39, 1.43, 1.47, and 1.51 using (c) GP and (d) ODT-DART.

## *4. Data summary*

The specifications about the demonstrated simulations and experiments are summarized in Table 2. The figure includes the essential information about the optical configuration and processing time. The processing time for ODT-DART is comparable to 100 iterations of GP.

**Table 2. Summarized specifications of data in simulations and experiments.**

| Data Voxel size | Beam Modulator | λ (nm) | M | Pixel (µm) | $N_{illum}$ | $NA_{obj}$ | $NA_{scan}$ | GP | | ODT-DART | |
|---|---|---|---|---|---|---|---|---|---|---|---|
| | | | | | | | | Noise | Time (s) | Initial ns | Time (s) |
| Phantom Silica bead 300×300×300 | Simulation (Spiral scan) | 532 | 400 | 17 | 201 | 1.4 | 0.9 | Off | 340.48 | 1.46 | 343.59 |
| | | | | | | | | On | 348.56 | | 336.11 |
| Phantom encapsulating 300×300×300 | | | | | 201 | 1.4 | 0.9 | Off | 341.92 | 1.4, 1.46 | 339.35 |
| | | | | | | | | On | 341.62 | | 353.85 |
| Phantom Shepp-Logan 300×300×300 | | | | | 201 | 1.4 | 1.2 | Off | 341.98 | 1.39, 1.43, 1.47, 1.51 | 380.61 |
| | | | | | | | | On | 347.24 | | 380.39 |
| Phantom 200 nm bead 512×512×512 | | | | | 201 | 1.4 | 1.2 | Off | 1747.9 | 1.46 | 1718.5 |
| Experiment Silica bead 300×300×300 | DMD (Spiral scan) | | 60 | 4.8 | 73 | 1.4 | 0.9 | Experiment | 243.31 | 1.42 | 165.24 |
| Experiment RBC 300×300×300 | DMD (Spiral scan) | | | | 117 | 1.4 | 0.9 | | 230.78 | 1.4 | 163.49 |
| Experiment Water drop 300×300×300 | Galvo (Spiral scan) | | 400 | 17 | 60 | 1 | 0.7 | | 267.76 | 1.34 | 154.98 |
| Experiment Trophozoite 300×300×300 | Galvo (Spiral scan) | 633 | 210 | 17 | 441 | 1.4 | 1.23 | | 237.72 | 1.4, 1.5 | 179.16 |

## 5. Reconstruction of 3D complex RI using ODT-DART

We also demonstrate that ODT-DART can precisely reconstruct both the real and the imaginary parts of complex RI maps. Figure 8 presents both the simulation and experimental results, in which samples have both real and imaginary RI values (extinction coefficient; $\kappa$).

Figures 8(a)–(b) present the simulation results of GP and ODT-DART for a 5-µm-diameter weakly absorbing phantom microsphere ($n_s = 1.461 + 0.01i$). As in Fig. 8(a), GP reconstructed inhomogeneous artifacts in the $\kappa$ distribution. The artifact is more severe in the presence of noises, which causes noises throughout the reconstructed $\kappa$ distribution. On the contrary, figure 8(b) shows that ODT-DART precisely reconstructed a homogeneous weakly absorbing sphere ($\kappa = 0.010$), even in the presence of the noise. The reconstruction accuracy of ODT-DART is further supported by the MSE analysis. Whereas the MSEs of GP were 1.10 and 29.34 ppm for the absence and presence of noises, the MSEs of ODT-DART were 0.862 and 1.353 ppm, respectively.

We also reconstructed the $\kappa$ distribution in experiment on the Pf-RBC using GP and ODT-DART, as shown in Fig. 8(c). Both GP and ODT-DART could localize the highly absorptive hemozoin using $\kappa$ distributions. The measured $\kappa$ of the hemozoin using ODT-DART was 0.036. In contrast, the cytoplasm had negligibly small $\kappa$ values in both GP and ODT-DART. However, whereas GP could not distinguish the RBC structure due to the missing cone and experimental noises, ODT-DART could distinguish the $\kappa$ value of the cytoplasm, which was estimated as 0.003.

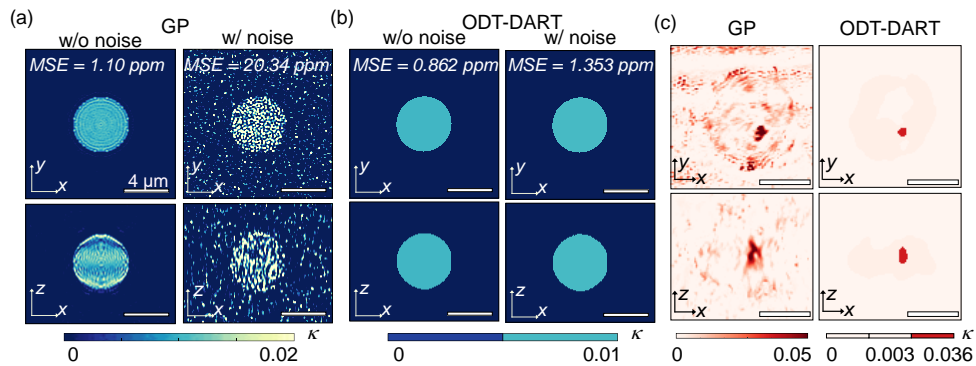

Fig. 8. Reconstruction of extinction coefficient $\kappa$ using ODT-DART. (a-b) cross-sectional images of the $\kappa$ maps of simulated 5-µm-diameter microspheres ($n_s = 1.461 + 0.01i$) reconstructed by (a) GP and (b) ODT-DART. (c) Cross sectional $\kappa$ maps of a Pf-RBC in Fig. 4 reconstructed by GP and ODT-DART.

**Funding**

This work was supported by KAIST, BK21+ program, Tomocube, and the National Research Foundation of Korea (2015R1A3A2066550, 2014M3C1A3052567, 2014K1A3A1A09-063027).

**Acknowledgement**

We thank Dr. Kyoohyun Kim and Mr. Jaehwang Jung for helpful discussion regarding the experiment and simulation analyses, and Mr. Sangyun Lee for the image visualization. Mr. Shin, Mr. Lee, and Prof. Park have financial interests in Tomocube Inc., a company that commercializes optical diffraction tomography and quantitative phase imaging instruments and is one of the sponsors of the work.